\documentclass[a4paper,11pt]{article}
\pdfoutput=1 

\usepackage{jinstpub} 

\title{\boldmath New High-Precision Drift-Tube Detectors for the ATLAS Muon Spectrometer}

\author[a,1]{H. Kroha,\note{Corresponding author.}}
\author[b]{R. Fakhrutdinov}
\author[a]{O. Kortner}
\author[b]{A. Kozhin}
\author[a]{K. Schmidt-Sommerfeld}
\author{and E. Takasugi}

\affiliation[a]{Max-Planck-Institute for Physics, F\"ohringer Ring 6, 80805 Munich, Germany}
\affiliation[b]{Institute for High Energy Physics, Science Square 1, Protvino, 142281 Russia}

\emailAdd{kroha@mppmu.mpg.de}

\abstract{Small-diameter muon drift tube (sMDT) detectors have been developed for upgrades of the ATLAS muon spectrometer.
With a tube diameter of 15~mm, they provide an about an order of magnitude higher rate capability than the present ATLAS muon tracking detectors, 
the MDT chambers with 30~mm tube diameter. The drift-tube design and the construction methods have been optimised for mass production and 
allow for complex shapes required for maximising the acceptance. A record sense wire positioning accuracy of $5~\mu$m has been achieved with
the new design. 14 new sMDT chambers are already operational in ATLAS, further 16 are under construction for installation in the 2019-2020 LHC shutdown.
For the upgrade of the barrel muon spectrometer for High-Luminosity LHC, 96 sMDT chambers will be contructed between 2020 and 2024.}


\begin{document}
\maketitle
\flushbottom

\section{Introduction}

\begin{figure}[htb]

\hspace{-6mm}
\begin{tabular}{lr}
\includegraphics[width=0.50\columnwidth]{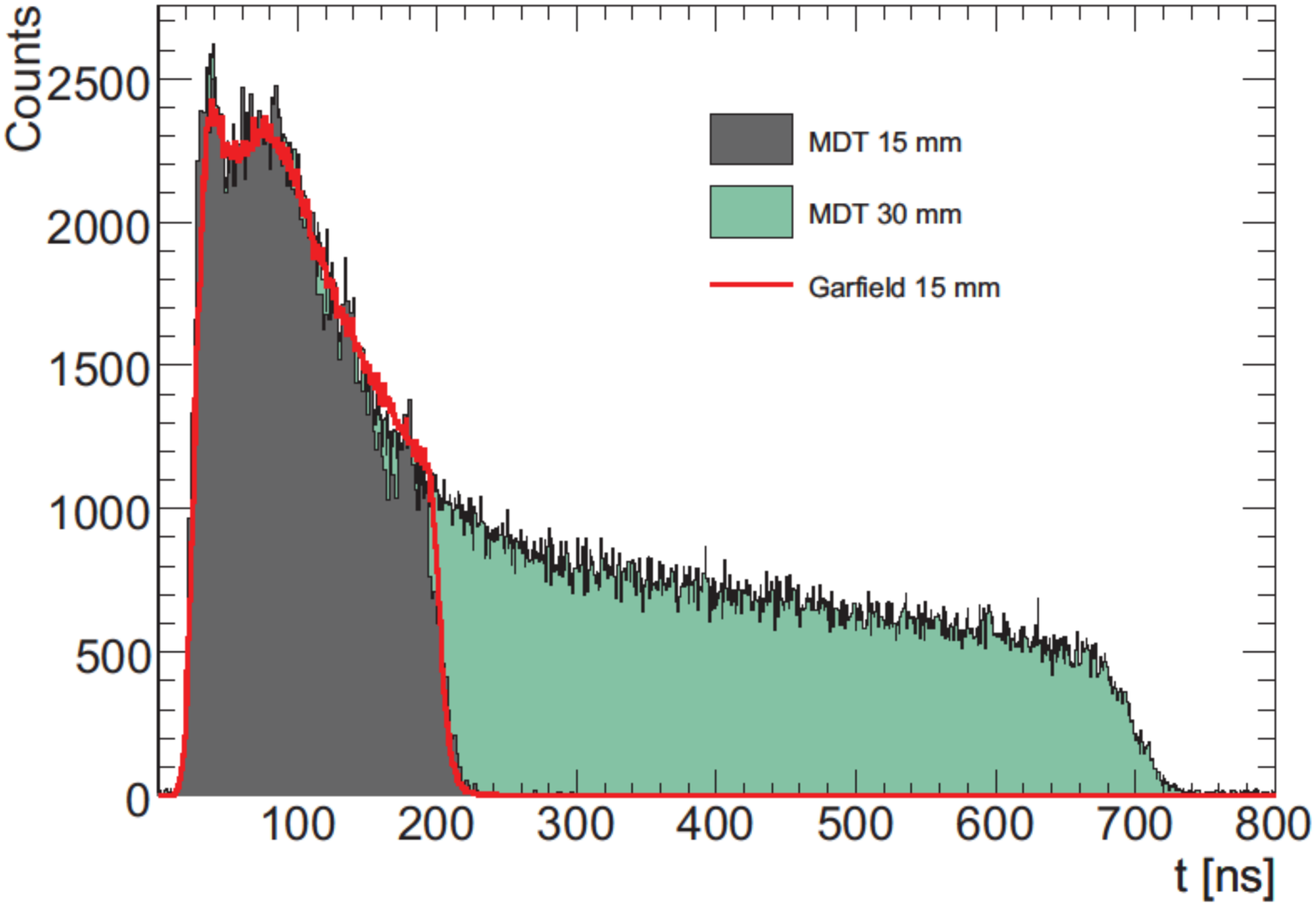} 
&

\includegraphics[width=0.50\columnwidth]{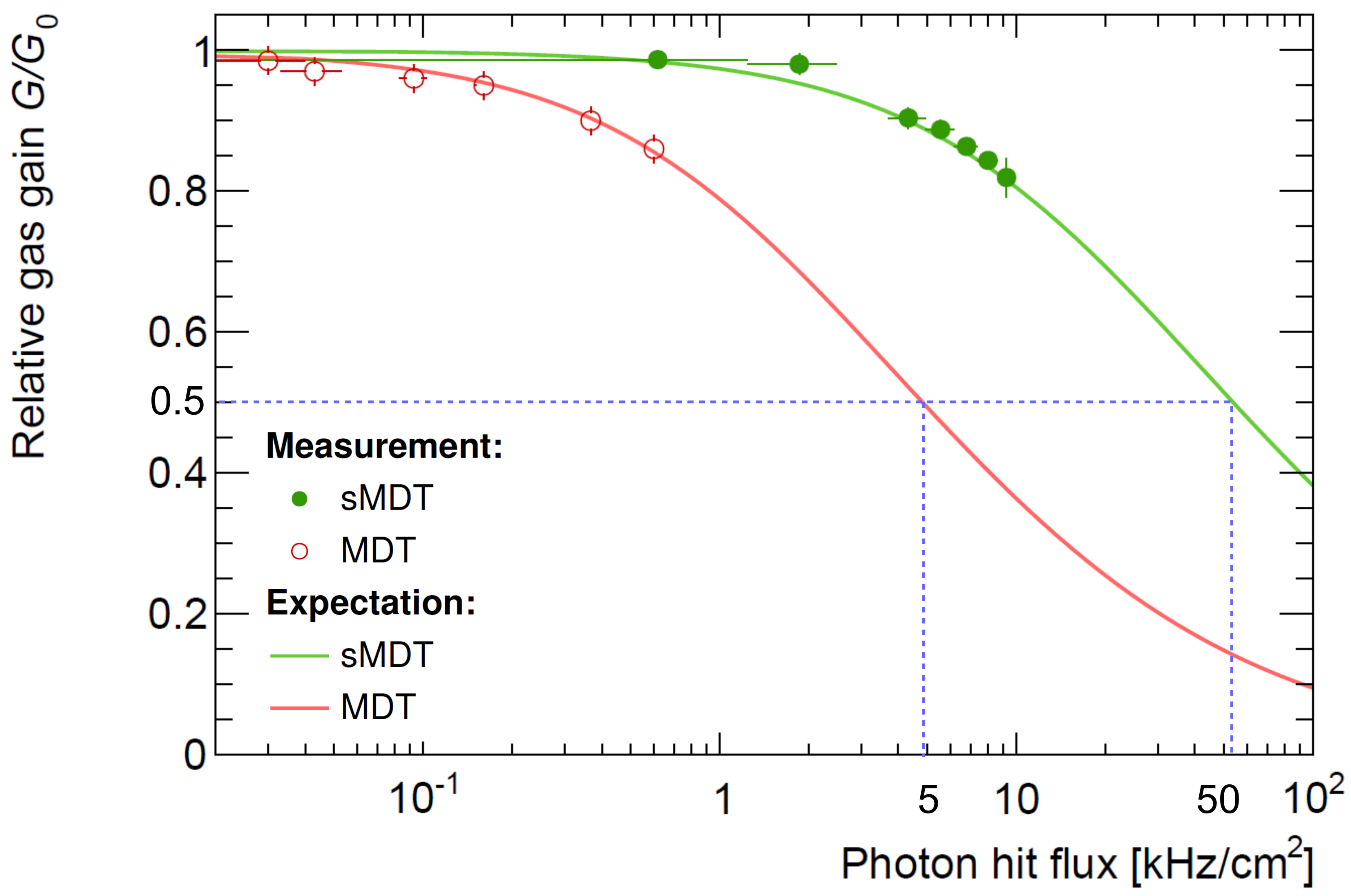} 
\\

\end{tabular}

\caption{Left: Drift time spectra of MDT (green) and sMDT tubes (grey) together with the prediction of a GARFIELD simulation for sMDT tubes (red line)~\cite{sMDTperf1}.
Right: Measurements of the gas gain of MDT and sMDT tubes relative to the nominal gas gain G$_0 =20000$ 
as a function of the $\gamma$ background rate at the Gamma Irradiation Facility at CERN compared to predictions based on the Diethorn formula~\cite{sMDTperf2}.}
\label{fig:spectrum_gasgain}

\end{figure}

The ATLAS Monitored drift tube (MDT) chambers~\cite{ATLASpaper} provide reliable muon tracking with excellent spatial resolution and high tracking efficiency 
independent of the track incident angle.
Small-diameter muon drift tube (sMDT) chambers with a tube diameter of 15~mm, i.e.\ half of the tube diameter of the MDT chambers, 
have been developed to cope with the higher background irradiation rates at High-Luminnosity LHC (HL-LHC) and future hadron colliders and to fit into small available spaces 
as it is necessary for the upgrades of the ATLAS muon spectrometer. At the same time, the chamber construction methods have been optimised for mass production with significant savings
in component cost, construction time and manpower compared to the ATLAS MDT chambers while providing the same reliability and mechanical robustness and 
even higher sense wire positioning accuracy. For the ATLAS precision muon tracking detectors a wire positioning accuracy of $20~\mu$m (rms) is required.
Standard aluminium tubes are used, with a wall thickness of 0.4 mm like for the MDT chambers.
The sMDT chambers are operated in ATLAS with the same gas mixture, gas pressure and gas gain
as the MDT chambers. Table~\ref{tab:parameters} shows a comparison of the MDT and sMDT operating parameters.
The drift time spectra are shown in the left-hand part of figure~\ref{fig:spectrum_gasgain}. The maximum drift time
of the sMDT tubes is only 175~ns compared to about 720~ns of the MDT chambers leading, together with the twice smaller cross section exposed to the radiation, 
to about 8 times lower occupancy and a linear space-to-drift time relationship with the standard MDT drift gas Ar:CO$_2$ (93:7) at 3 bar pressure.

\begin{table}[ht]

\begin{center}
\caption{Material and operating parameters of ATLAS sMDT chambers~\cite{sMDTperf1} compared to the MDT chambers~\cite{ATLASpaper}. 
500 Hz/cm$^2$ and 200~kHz/tube are the maximum background rates expected in the ATLAS muon drift-tube chambers at HL-LHC.}
\label{tab:parameters}
\vspace{2mm}

\begin{tabular}{|l|c|c|}

\hline
Type & MDT & sMDT \\
\hline\hline
Tube material & Aluminium & Aluminium \\
              & Aluman100 & AW 6060-T6/ AlMgSi \\
Tube inner$\&$outer surface & & Surtec 650 chromatisation \\
Tube outer diameter & 29.970~mm & 15.000~mm \\
Tube wall thickness & 0.4~mm & 0.4~mm \\
Wire material & W-Re (97:3) & W-Re (97:3) \\
Wire diameter & $50~\mu$m & $50~\mu$m \\
with gold plating, thickness & $3\%$ & $3\%$ \\  
Wire resistance/m & $44~\Omega /$m & $44~\Omega /$m \\
Wire pitch & 30.035~mm & 15.099~mm \\
Wire tension & $350\pm 15$~g & $350\pm 15$~g \\
\hline
Gas mixture & Ar:CO$_2$ (93:7) &  Ar:CO$_2$ (93:7) \\
Gas pressure & 3 bar (abs.) & 3 bar (abs.) \\
Gas gain & $2\cdot 10^4$ & $2\cdot 10^4$ \\
Wire potential & 3080~V & 2730~V \\
Maximum drift time & 720~ns & 175~ns \\
\hline
Average tube spatial resolution & $83~\mu$m & $106~\mu$m \\
without backgr.\ irradiation & & \\
Average tube spatial resolution & $160~\mu$m & $110~\mu$m \\
at 500 Hz/cm$^2$ backgr.\ rate & & \\
Drift tube muon efficiency  & $95\%$ & $94\%$ \\ 
without backgr.\ irradiation & & \\
Drift tube muon efficiency  & $80\%$ & $90\%$ \\ 
at 200 kHZ/tube backgr.\ rate & & \\
Wire positioning accuracy & $20~\mu$m (rms) & $10~\mu$m (rms) \\
\hline
\end{tabular}

\end{center}

\end{table}

\begin{figure}[htb]
\begin{tabular}{lr}

\includegraphics[width=0.54\columnwidth]{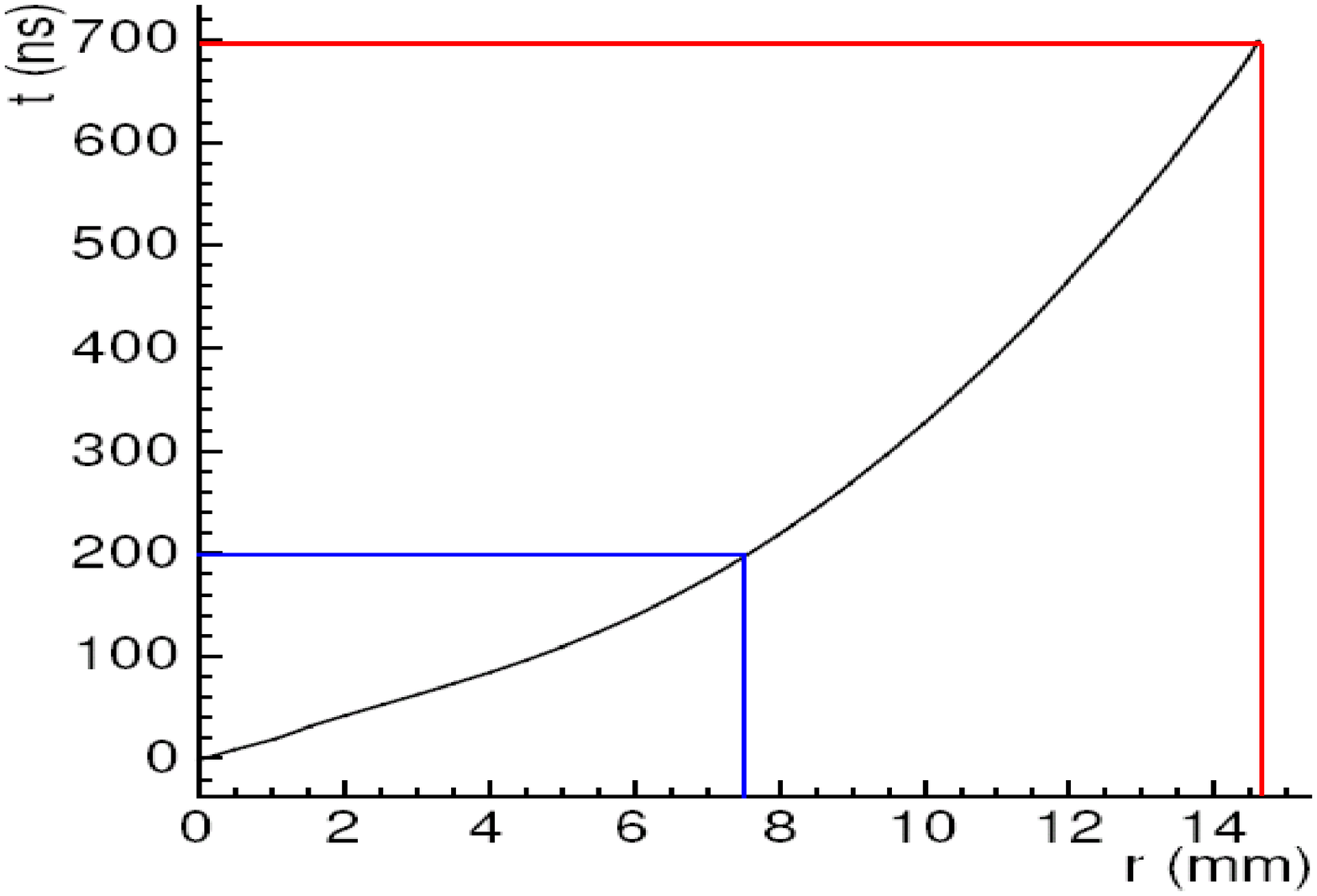} 
&

\hspace{-10mm}
\includegraphics[width=0.50\columnwidth]{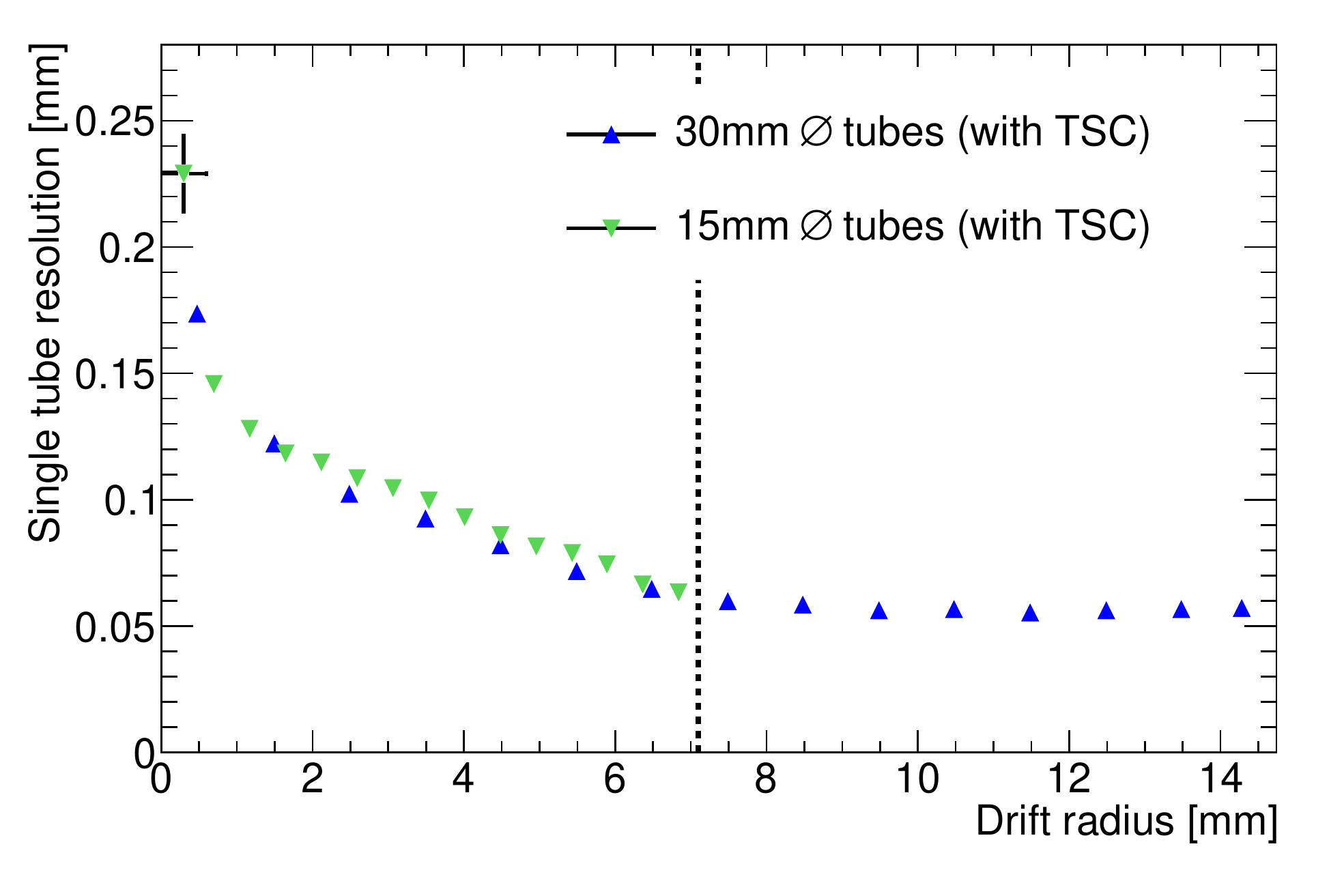} 
\\

\end{tabular}

\caption{Left: Track radius-to-drift time relationship of MDT drift tubes. The part with drift radii r below 7.5~mm, relevant for the sMDT tubes, is linear to good approximation.
Right: Spatial resolution after time slewing corrections (TSC) as a function of the drift radius for MDT and sMDT drift tubes measured under the same operating conditions 
in the H8 muon beam at CERN without background irradiation~\cite{sMDTperf2}.
As expected, the results for 15 and 30~mm diameter tubes are in good agreement for drift radii below 7.5~mm.}
\label{fig:rtrelation_resolr}

\end{figure}

\begin{figure}[htb]
\centering
\includegraphics[width=0.7\columnwidth]{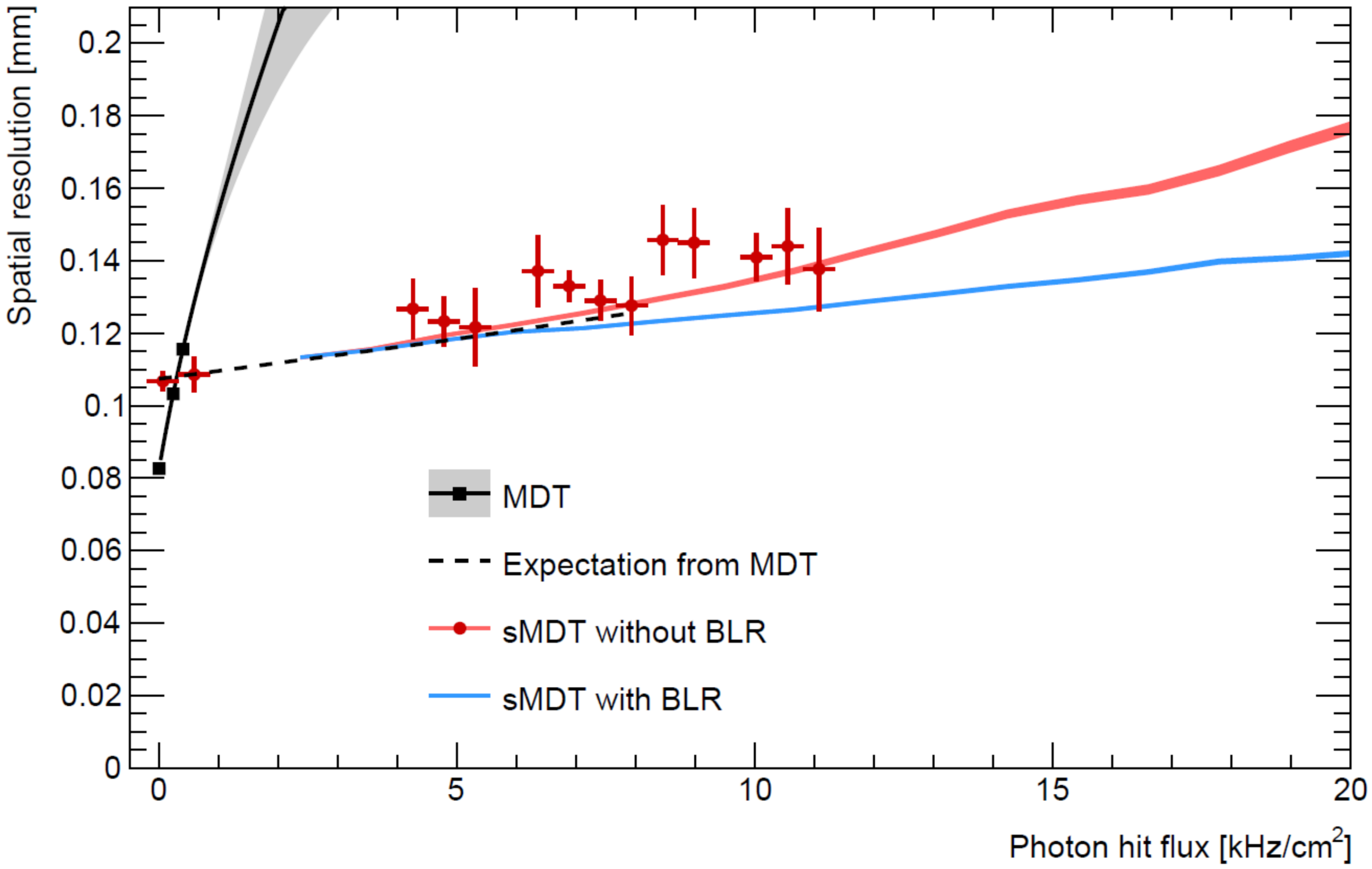}
\caption{Average spatial resolution of MDT and sMDT drift tubes measured at the Gamma Irradiation Facility at CERN as a function of the $\gamma$ background rate 
using standard MDT readout electronics with bipolar shaping. The same front-end electronics scheme and parameters will be used for MDT and sMDT chambers at HL-LHC.
Further improvement of the sMDT drift tube spatial resolution at high background rates and space charge densities can be achieved by employing additional fast baseline restoration (BLR)
in order to suppress signal pile-up effects (blue curve)~\cite{blr} which is not needed for operation at HL-LHC.}
\label{fig:resolution}
\end{figure}

\begin{figure}[htb]
\centering
\includegraphics[width=0.7\columnwidth]{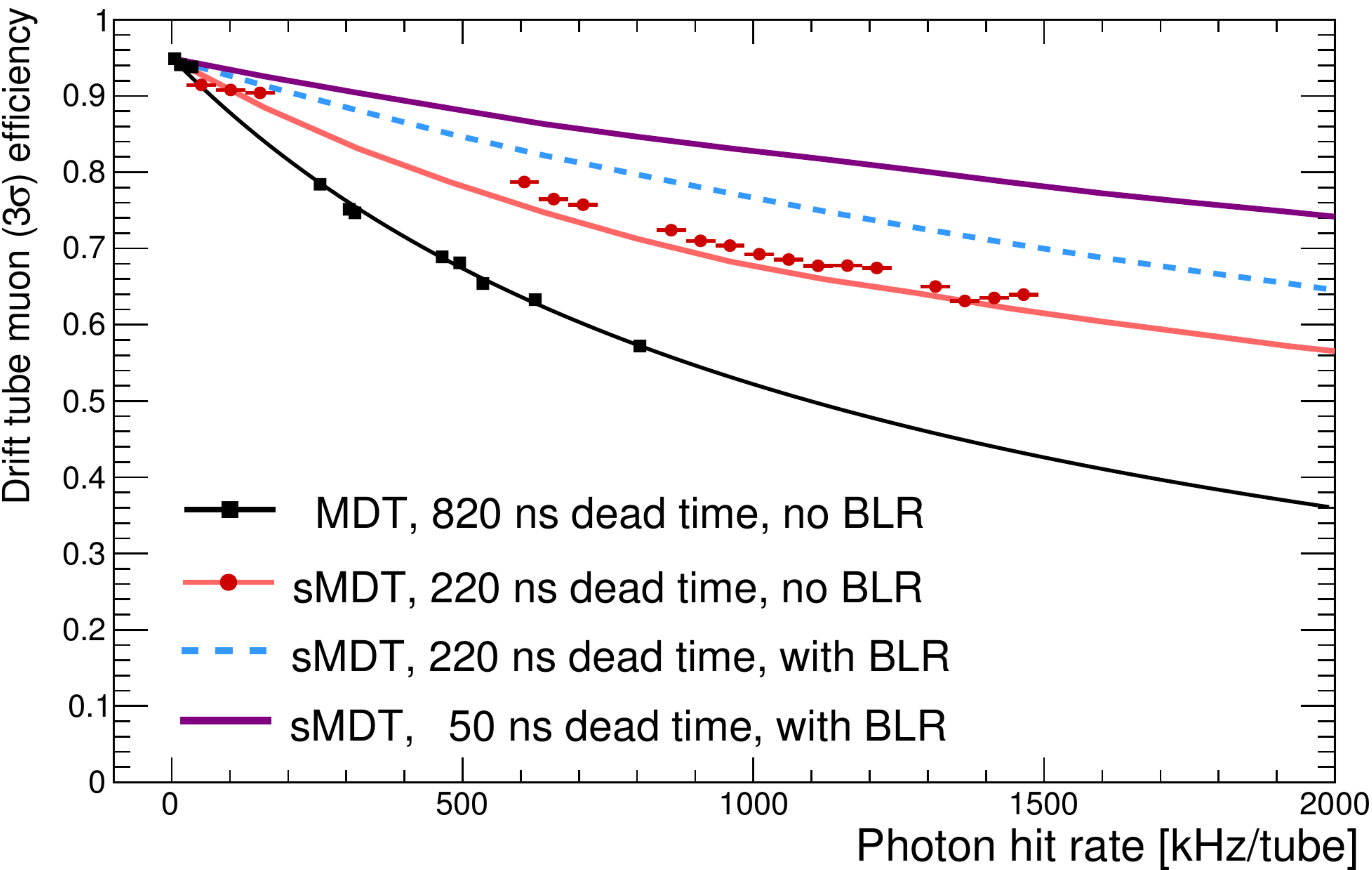}
\caption{Muon detection efficiencies of  MDT and sMDT drift tubes with corresponding electronics deadtime settings (see text) measured at the Gamma Irradiation facility at CERN
as a function of the $\gamma$ background counting rate using standard MDT readout electronics with bipolar shaping as planned also for operation at HL-LHC.   
The efficiency is defined as the probablity to find a hit on the extrapolated track within $3\sigma$ of the drift tube spatial resolution ($3\sigma$ efficiency). 
The measurement results agree well with the expectations from detailed simulations of detector and electronics response. 
Further improvements of the muon efficiency of the sMDT drift tubes at high background counting rates are possible by employing fast baseline restoration (BLR) 
in order to suppress signal pile-up effects (blue dashed curve)~\cite{blr} which is not needed for operation at HL-LHC.}
\label{fig:efficiency}
\end{figure}

\begin{figure}[ht]
\vspace{-18mm}
	\includegraphics[width=1.15\textwidth]{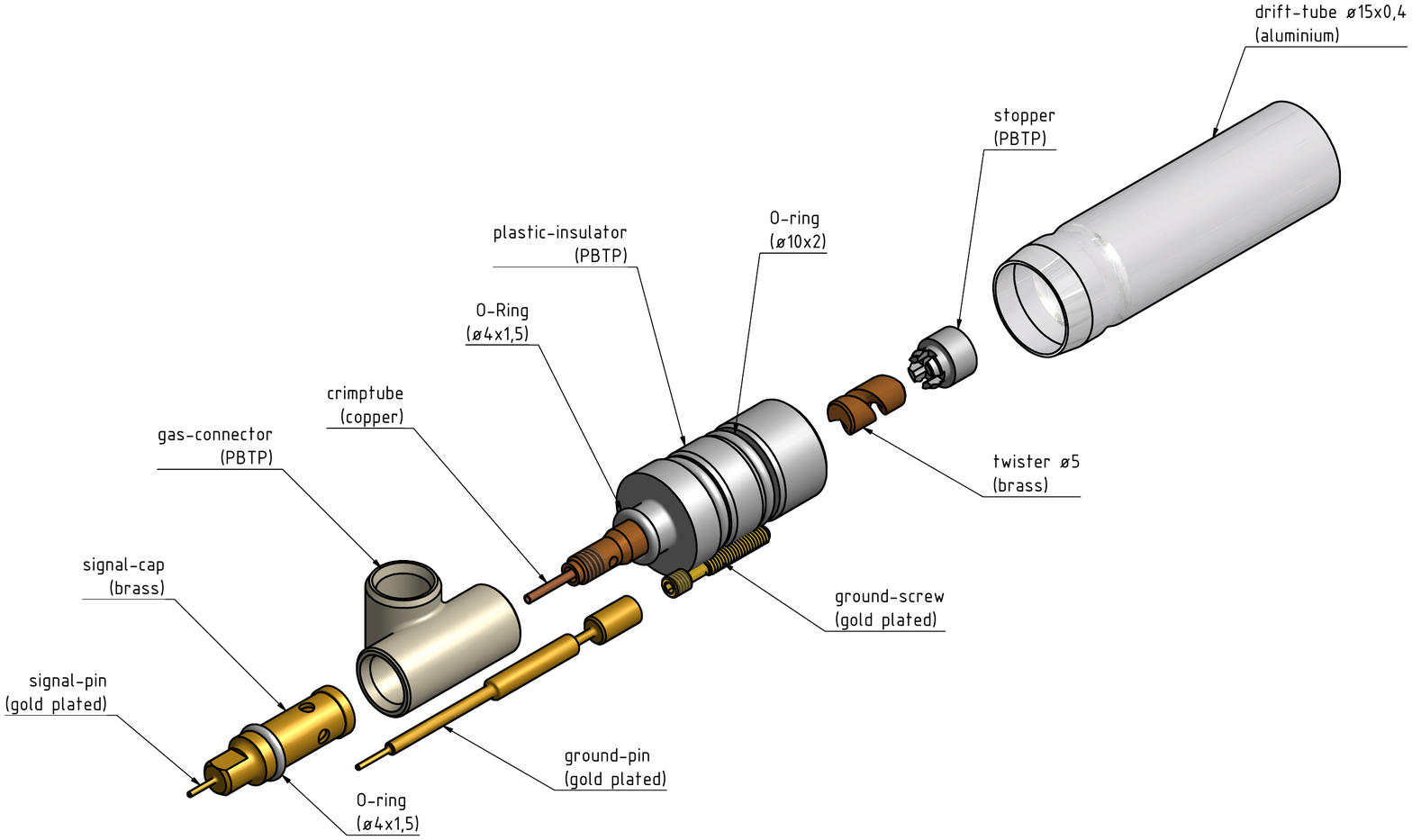}
	\caption{Exploded view of an sMDT endplug with interfaces for precise wire positioning and measurement, for gas and high-voltage supplies and for readout 
        electronics~\cite{sMDTprototype}.}
	\label{fig:endplug1}

\vspace{2mm}

	\centering
	\includegraphics[angle=-90,width=0.7\textwidth]{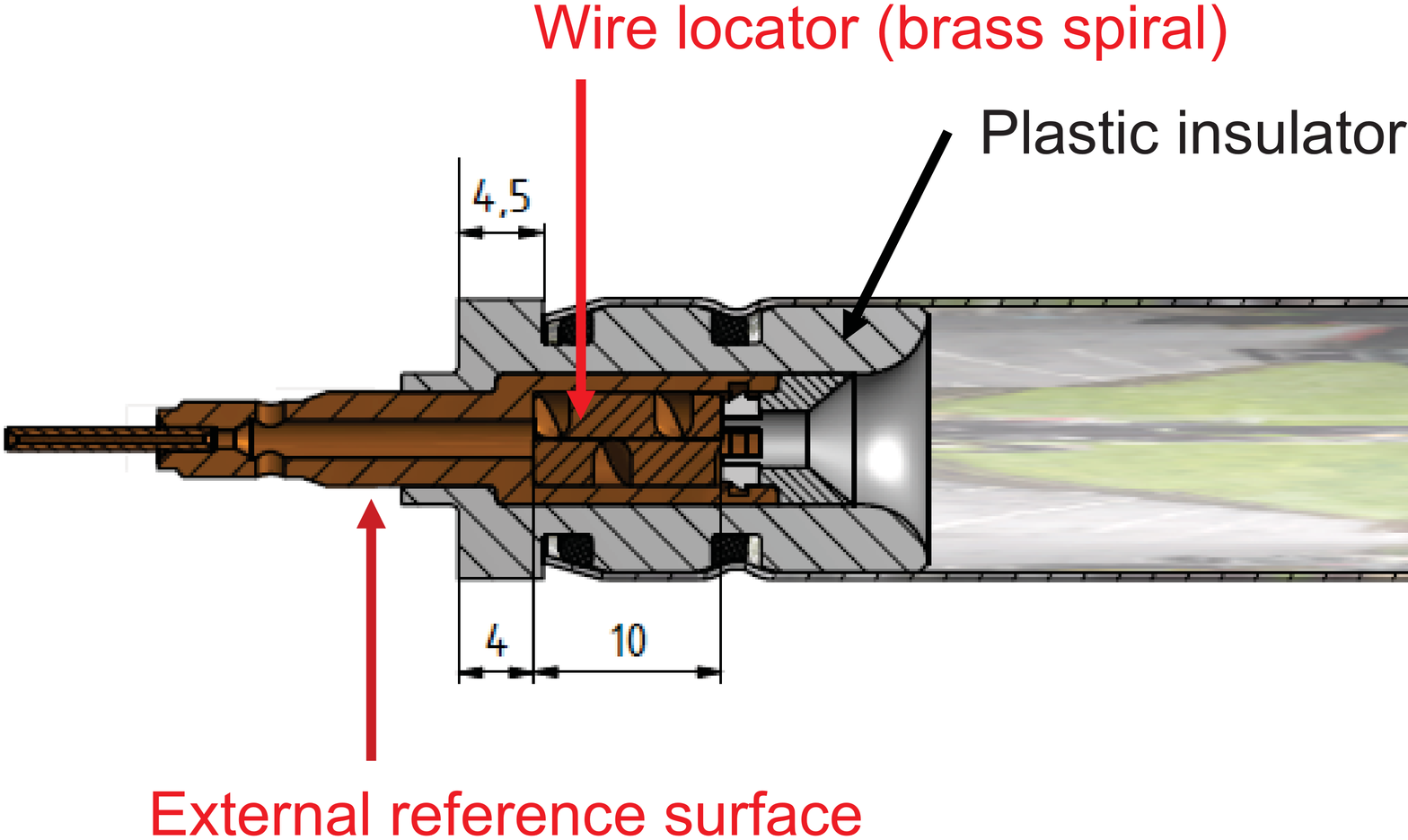}
	\caption{Cross section of an sMDT endplug with internal wire locator and external reference surface for tube and wire positioning during construction and for 
	wire position measurement~\cite{BMGconstr}.}
	\label{fig:endplug2}
\end{figure}




\begin{figure}[th]
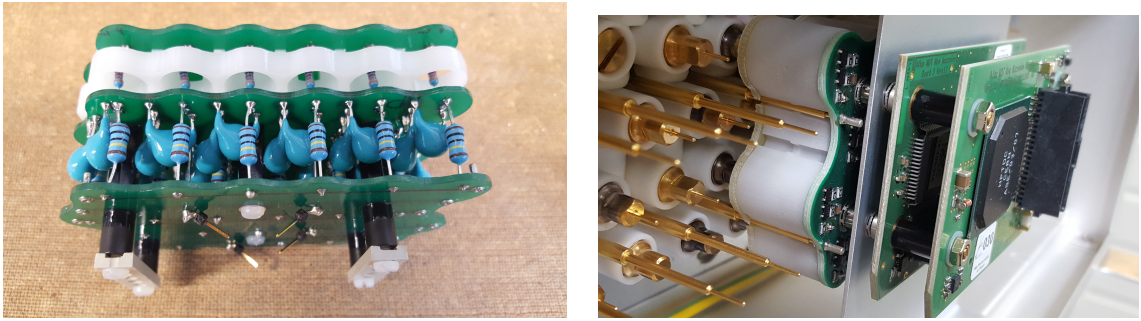

\begin{tabular}{lr}
	\includegraphics[width=0.49\textwidth]{sMDT_HVHH_BMG.pdf} &
	
	\includegraphics[width=0.47\textwidth]{sMDT_mezzanine_card_BMG.pdf} \\
	
\end{tabular}	
	\caption{Left: sMDT high-voltage distribution (hedgehog) board. Right: sMDT signal distribution board with stacked active readout electronics (mezzanine) card, 
	carrying three ASD chips in the middle layer and the TDC chip with programming FPGA in the top layer, mounted on the gold-plated signal and ground pins of the drift tubes.
	The termination resistors on the high-voltage side and the coupling capacitors on the readout side are encapsulated in injection molded white plastic containers 
	in order to ensure HV stability.}
	\label{fig:sMDT_electronics}
\end{figure}


\begin{figure}[th]
\centering
	\includegraphics[width=0.6\textwidth]{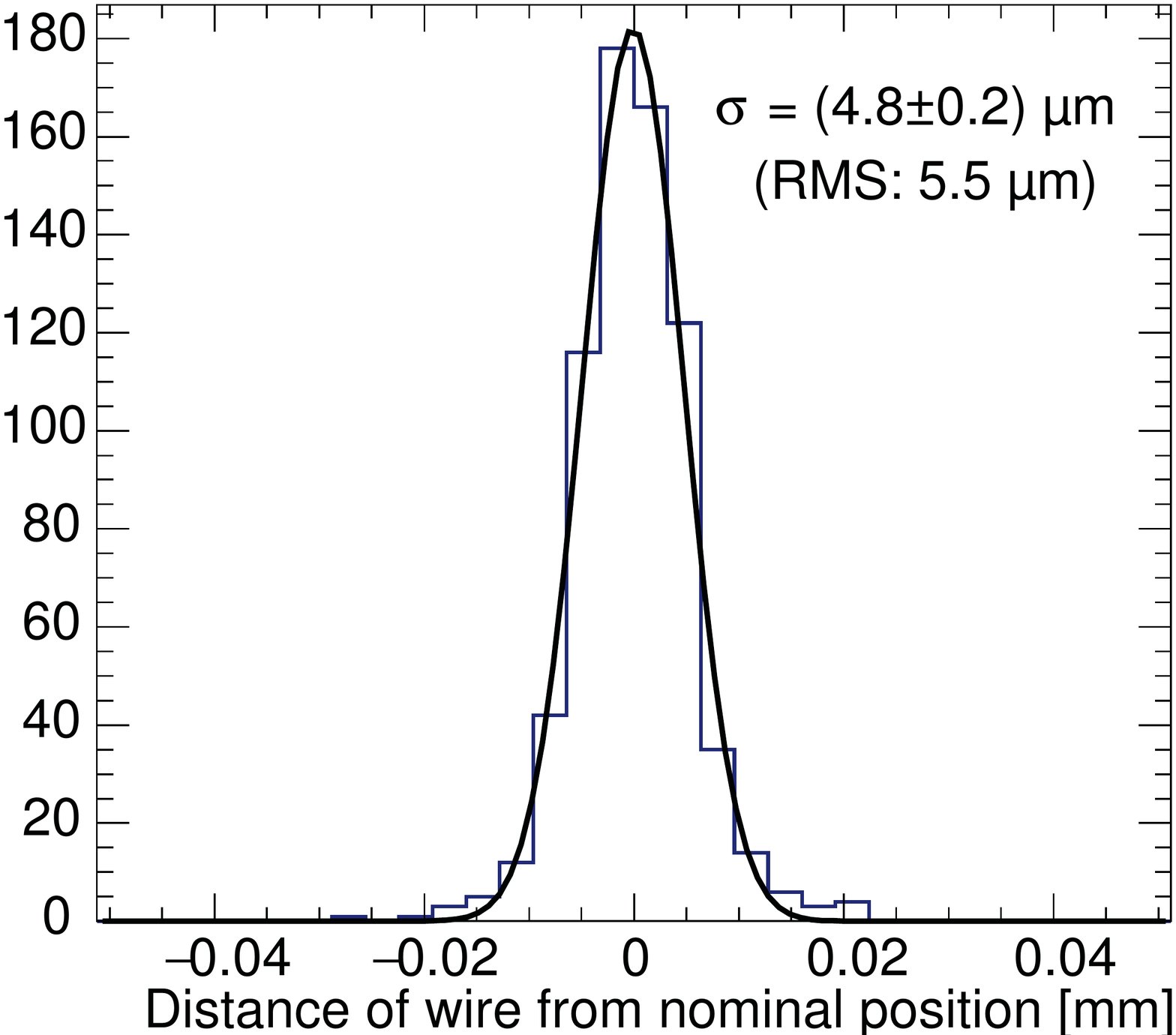}
	\caption{Residuals of the sense wire positions measured at both ends of a BMG sMDT chamber with 356 tubes with respect to the nominal wire grid. 
	The width of the distribution includes the accuracy of the coordinate measureing machine of about $2~\mu$m~\cite{BMGconstr}.}  	
	\label{fig:residuals}
\end{figure}

\begin{figure}[ht]
\begin{tabular}{lr}
	\includegraphics[width=0.48\textwidth]{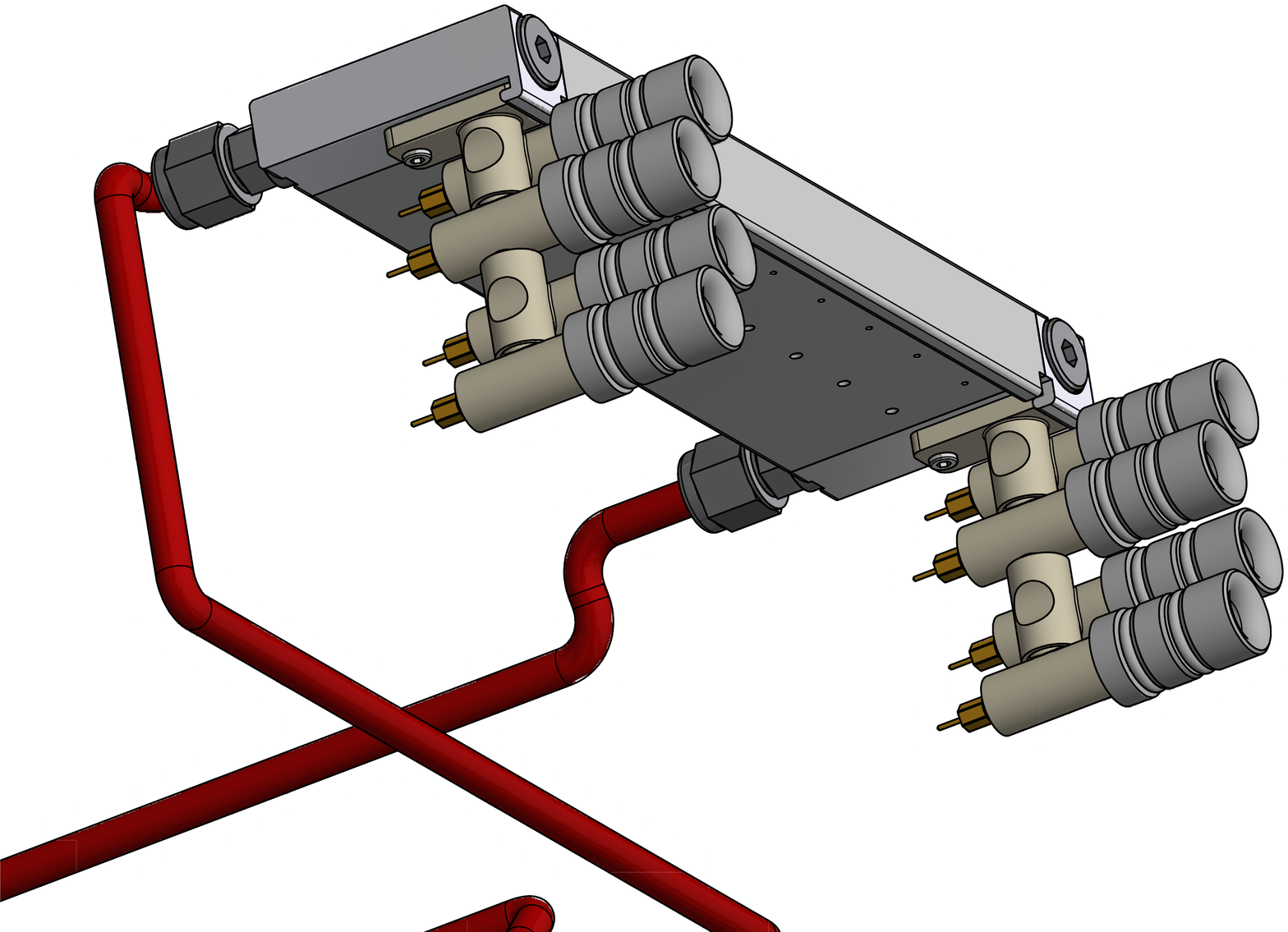} &

	\includegraphics[width=0.48\textwidth]{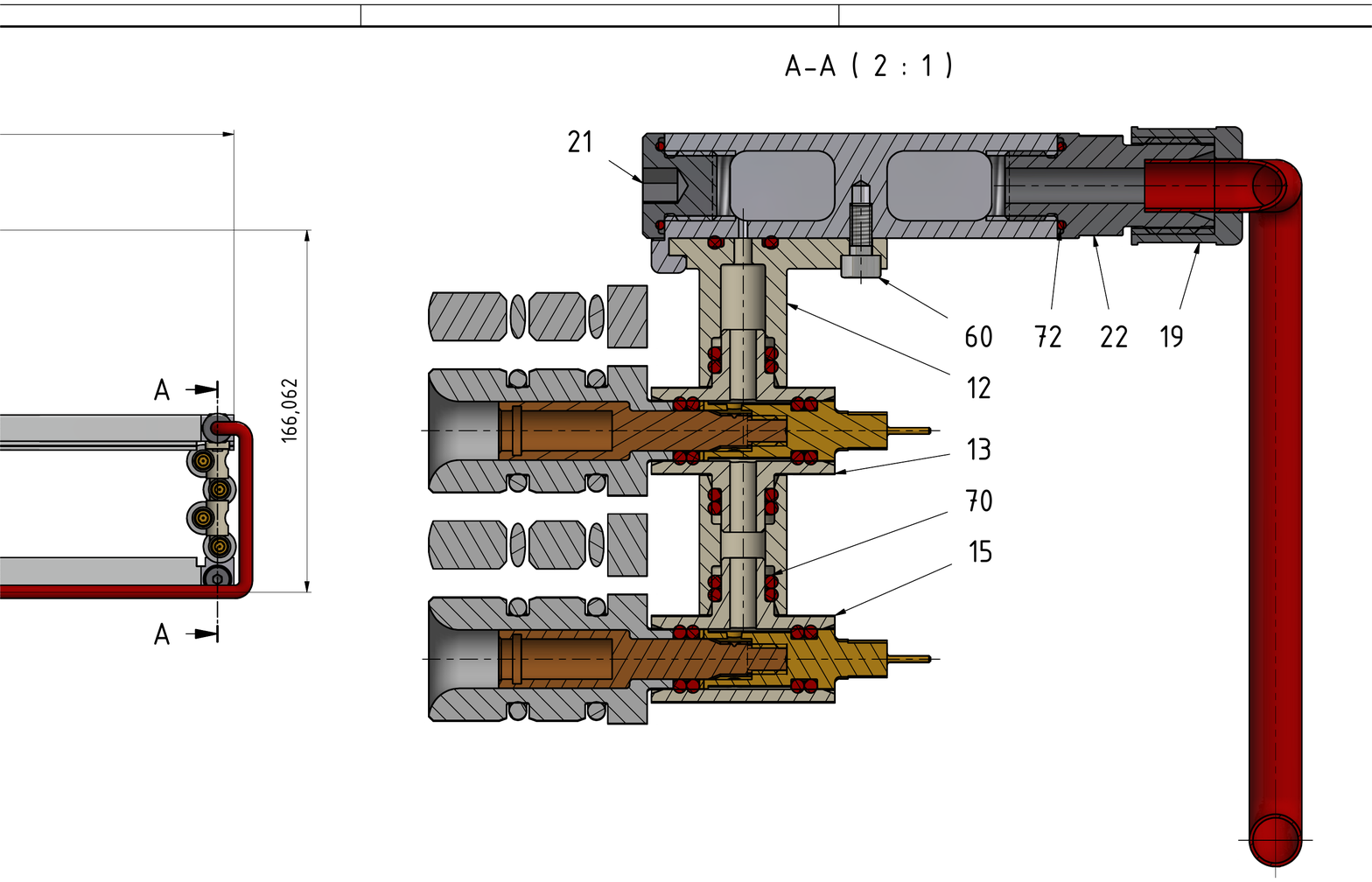} \\	
	
\end{tabular}	
	\caption{Schematics of the gas distribution system of the BIS~7/8 and BIS~1-6 sMDT chambers (left: 3D model, right: cross section). 
	The principle is the same as for the BME and BMG chambers. It uses injection molded plastic gas connectors made of Crastin S600F20 (PBTB), 
	the same materials as used for the endplug plastic parts~\cite{BMGconstr} 
        (see Figure~\ref{fig:endplug1}), to connect the tubes in columns perpendiculat to the chamber plane to the 
        aluminium gas bars mounted along each multilayer on the HV and RO side.}  
	\label{fig:bis78_gassystem}
\end{figure}

\begin{figure}[ht]

\vspace{-30mm}


\includegraphics[width=1.1\columnwidth]{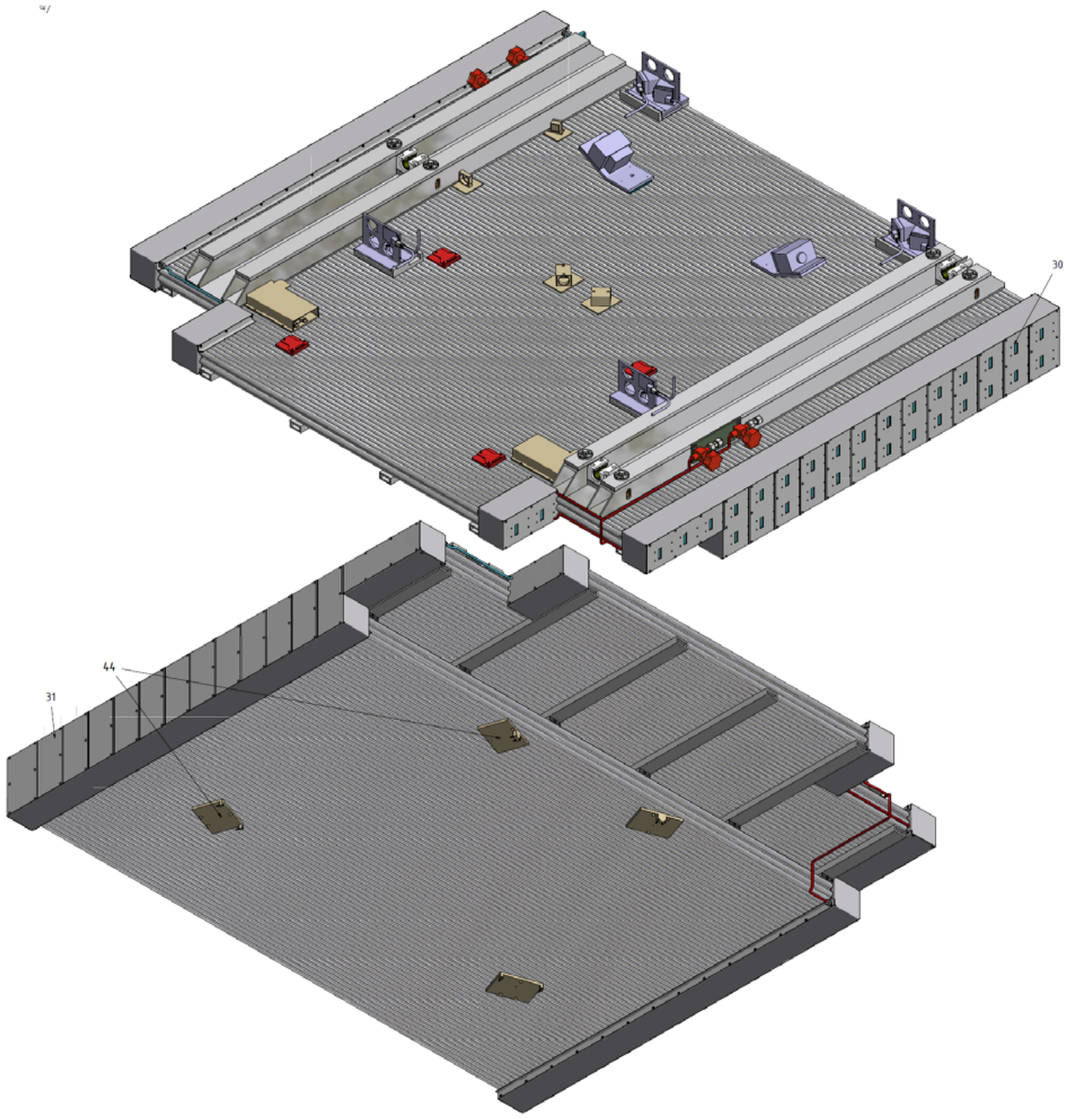} 


\vspace{-70mm}

\caption{Schematic view of a BIS sMDT chamber for upgrade of the ATLAS muon spectrometer in 2019-2020. The alignment sensors at the top and the bottom of the
chamber are mounted on the outer drift tube layers with a precision of $20~\mu$m with respect to the sense wires during chamber assembly.}
\label{fig:bis78_4c_chamber}

\end{figure}

A full-scale sMDT prototype chamber of trapezoidal shape has been constructed and tested in the H8 muon beam and in the Gamma Irradiation Facility (GIF) 
at CERN in 2010~\cite{sMDTprototype}. The chamber has been operated in the ATLAS cavern in 2012. 
In 2014, two sMDT chambers~\cite{BMEop}, each with two integrated RPC chambers, have been installed in access shafts in the feet region of the ATLAS barrel muon spectrometer
(so-called BME chambers) and are in operation since the start of LHC run~2. In January 2017, 12 new sMDT chambers
have been installed inside the detector feet in the bottom sectors of the barrel muon spectrometer (so-called BMG chambers)~\cite{BMEop,BMGconstr,BMGproduction}
and are in operation for the new data taking in 2017. 

The construction of further 16 sMDT chambers with integrated triplet RPC trigger chambers (see figure~\ref{fig:bis78_4c_chamber}) has started. They will be installed under 
very tight spatial constraints 
on the toroid magnet coils at the ends of the inner barrel layers (so-called BIS chambers) in the long LHC shutdown in 2019-2020 in order to improve the trigger efficiency and the rate capability 
of the chambers in the transition regions between barrel and endcaps. They have rather complex shapes in order to maximise the acceptance in the overlap 
region between the barrel part the muon spectrometer and the inner endcap layer and can only be built with the assembly methods developed for the sMDT chambers.  
This upgrade of the muon spectrometer serves as pilot project for the complete replacement of the MDT chambers in the  
by sMDT-RPC chamber modules enhancing the rate capability of the tracking and trigger chambers by about an order of magnitude and increasing the barrel muon trigger efficiency 
and robustness for operation at HL-LHC. The installation of new triple thin-gap RPCs of only 5~cm thickness becomes possible only by replacing the BIS MDT chmabers
by sMDT chambers which have about half the height. 96 new BIS sMDT chambers will be constructed for this purpose in the years 2020-2024.

\section{Performance of the sMDT chambers}
The performance of MDT~\cite{MDTperf} and sMDT chambers~\cite{sMDTperf1} has been extensively studied at the Gamma Irradiation Facility at CERN using the existing ATLAS MDT readout electronics with 
bipolar shaping. For the (s)MDT amplifier-shaper-discriminator (ASD) chips at HL-LHC the same specifications will be used 
as for the  present system. The MDT chambers can be operated up to background rates of 500~Hz/cm$^2$ and 300~kHz per tube.  
At background rates above 500~Hz/cm$^2$, the gas gain drops by more than $20\%$  
(see figure~\ref{fig:spectrum_gasgain}, right) leading, together with the effect of space charge fluctuations, to rapid deterioration of the spatial resolution with increasing background flux. 
The limitations of the MDT chambers are overcome by using drift tubes with half the diameter of the ATLAS MDT tubes while leaving the operating parameters, 
Ar:CO$_2$ (93:7) gas mixture at 3 bar pressure and nominal gas gain of 20000 (for a wire potential with respect to the tube wall of 2730~V in sMDT tubes), unchanged~\cite{sMDTperf1}. 

As the space charge density inside the drift tubes is proportional to the third power of the tube radius, 15~mm diameter drift tubes show a significant gain drop only at 8 times higher 
background rates compared to 30~mm diameter drift tubes (see figure~\ref{fig:spectrum_gasgain}, right). 
At the same time, the deteriorating effect of space charge fluctuations on the spatial resolution is eliminated because the drift gas is linear to good approximation for drift radii below 7.5~mm 
(see figure~\ref{fig:rtrelation_resolr}, left). The radial dependence of the spatial resolution of 15 and 30~mm diameter drift tubes from measurements in the H8 muon beam at CERN without 
radiation background is shown in the right-hand part of figure~\ref{fig:rtrelation_resolr}~\cite{sMDTperf2}. Standard MDT time-slewing corrections are applied in both cases.  
Without irradiation and associated space charge effects and with time-slewing corrections, the average sMDT drift tube resolution is $106\pm 2~\mu$m 
compared to $83\pm 2~\mu$m for the MDTs~\cite{sMDTperf2}.
The dependence of the average spatial resolution of MDT and sMDT drift tubes on the $\gamma$ background rate is shown in figure~\ref{fig:resolution}. The spatial resolution deteriorates quickly
with increasing background flux for the MDTs while it is affected only little by space charge effects up to very high irradiation rates for the sMDTs. 

At the same background rate, the small-diameter drift tubes experience 8 times lower occupancy than the 30~mm diameter MDT tubes because of the 4 times shorter maximum drift time 
(see figure~\ref{fig:spectrum_gasgain})
and the twice smaller tube cross section exposed to the radiation. Because of the much shorter maximum drift time, the dead time of the MDT readout electronics 
(which for the MDTs is set to a nominal value of 820~ns, slightly above the maximum drift time, to prevent the detection of secondary ionization clusters) 
can be reduced to the minimum adjustable value of 220~ns, just above the maximum drift time of the sMDT tubes. 
In this way, the masking of muon hits by preceding background pulses is strongly reduced increasing the muon detection efficiency defined as the probability to find a hit on the extrapolated 
muon track within 3 times the drift tube resolution ($3\sigma$ efficiency). Figure~\ref{fig:efficiency} shows the improvement of the $3\sigma$ efficiency of sMDT tubes at high 
background counting rates compared to the MDT tubes. 
Muon track segment reconstruction efficiencies of almost $100\%$ and a spatial resolution of better than $30~\mu$m are achieved with 8-layer sMDT chambers at the maximum background rates 
expected at HL-LHC.

\section{Drift tube design and fabrication}
The sMDT chamber design and construction procedures have been optimized for mass production while they provide highest mechanical accuracy in the sense wire positioning.
Standard industrial aluminium tubes with 15~mm outer diameter and a wall thickness of 0.4~mm are used. The tubes are chromatised on the in- and outside for the cleaning purposes 
and reliable electrical ground contact. The ground pins are screwed into the holes between adjacent tube triplets during the glueing of the tube layers (see figure~\ref{fig:endplug1}).
The drift tube design and fabrication procedures are the same as used for the construction of the BMG sMDT chambers in 2016~\cite{BMGconstr, BMGproduction}.

The drift tubes are assembled using a semi-automated wiring station in a temperature-controlled clean room. 
The endplugs are inserted into the tubes and the sense wires fed through the tubes and the endplugs 
by means of air flow without manual contact. Afterwards the endplugs are fixed and and the tubes gas sealed by swaging. Finally,
the wires are fixed in copper crimp tubelets inserted in the endplug central pins after tensioning them to $350\pm 15$~g, 
corresponding to a gravitational sag of only $17\pm 1~\mu$m (absolute tolerances) for 1~m long tubes, including overtensioning to 430~g for 10~s.
The wires are positioned at each tube end with a few micron precision with respect to a cylindrical external reference surface
on the central brass insert of the endplug which also holds the spiral shaped wire locator on the inside of the tube. 
The drift tubes are sealed with the endplugs using two O-rings per endplug and mechanical swaging of the tube walls. For the injection molded endplug insulators 
and gas connectors for the individual tubes, plastic materials with minimum outgassing have been selected which are also immune against cracking.

Only materials already certified for the ATLAS MDT chambers are used for sMDT drift tubes and their gas connections in order to prevent ageing.
No outgassing of the plastic materials of endplugs (PBTP Crastin LW9330, reinforced with 30\% glass fiber) and gas connectors (PBTP Crastin S600F20, unreinforced) has been observed.
The sMDT tubes, including the plastic material of the endplugs, have been irradiated with a 200 MBq ${}^{90}$Sr 
source over a period of 4 months with a total charge accumulation on the sense wire of 9~C/cm without any sign of aging~\cite{sMDTperf2,aging}.

Typical production rates of 100 tubes per day have been achieved with one assembly station operated by two technicians at an average failure rate of about $4\%$, which is
mostly due to occasional failures of the assembly devices.
During the production of the 4300 BMG drift tubes, the failure rate of the standard drift tube quality tests of wire tension ($350\pm 15$~g), 
gas leak rate ($<10^{-8}$~bar~l/s) and leakage current ($< 2$~nA/m) at the nominal operating voltage of 2730~V was only $2\%$, mostly due to too high dark currents
under high voltage.

\section{sMDT chamber construction and test}
After passing the quality assurance tests, the drift tubes are assembled to chambers in a climatised clean room by inserting the endplug reference 
surfaces into a grid of fitting bores in the assembly jigs 
at each chamber end which define the wire positions with an accuracy of better than $5~\mu$m 
and glueing them together and to the spacer and support frame 
using an automated glue dispenser. 
A complete chamber can be assembled within two working days, including the precise mounting of the global alignment sensor platforms. 
The same two-component expoxy glues as for the MDT chamber construction are used, Araldite~2014 between the tube layers and DP~490 between multilayers and spacer  and support structures.
After the glueing of each new tube layer, ground connection screws are inserted into the triangular gaps between adjacent tube layers through holes in the jig, scratching the chromatised tube walls.
The gaps are filled with glue during the assembly of the next layer, fixing and encapsulating the ground screws. Conducting glue may be added in order to improve the conductivity 
of the ground connection if necessary.  
After mounting of the gas connections, ground pins connecting to the readout and high-voltage distribution boards are screwed onto the ground screws  
(see figure~\ref{fig:sMDT_electronics}).

Like the BME sMDT chambers, but in contrast to the BMG chambers, the BIS7/8 and BIS~1-6 sMDT chambers will have in-plane alignment monitoring systems. The longitudinal sag monitors of the 
in-plane alignment system of the BIS~7/8 chambers is rotated by 180$^\circ$ with respect to the standard orientation parallel to the tube direction in the MDT and also the BME and BIS~1-6 chambers 
in order to properly monitor potential deformations of the complex shaped chambers transverse to the tubes. Two diagonal straightness monitors measure torsions between the readout and 
high-voltage ends in all types of chambers. Like the MDT chambers, the BME and BIS sMDT chambers carry an optical alignment system monitoring the planarity of the chambers. The BMG and
BIS sMDT chambers carry in addition optical sensors for the alignment of the chambers with respect to neighboring chambers, which are mounted on the tube layers with $20~\mu$m positioning accuracy 
with resepct to the sense wires during chamber assembly (see figure~\ref{fig:bis78_4c_chamber}). 

After the glueing of the tube layers, the positions of the individual endplug reference surfaces and, thus, of the sense wires are measured at the two chamber ends 
with an automated coordinate measuring machine with a precision of about $2~\mu$m. 
The measurement was performed within 1-2 hours for every BME and BMG chamber and is 
planned as regular spot check during the BIS chamber serial production. 
In particular, the positions of the alignment sensor platforms with respect to the wire grid can be measured with a few micron accurracy.
Sense wire positioning accuracies of better than $10~\mu$m (rms) have been routinely achieved during BME and BMG chamber construction~\cite{BMEop,BMGconstr}. 
An ultimate wire positioning accuracy of $5~\mu$m (rms) has been achieved in the BMG chamber construction, which comes close the precision of the 
assembly jigs (see figure~\ref{fig:residuals}). All BMG sMDT chambers have a wire positioning accuracy of better than $10~\mu$m with an average of $7~\mu$m. 
After the measurement, the individual wire positions are known with $2~\mu$m accuracy.
 
After the wire position measurement, the parallel gas distribution system is mounted, consisting of modular injection molded plastic gas connectors connecting
tubes in columns perpendicular to the chamber plane to the chromatised aluminium gas distribution bars  
(see figures~\ref{fig:endplug1} and \ref{fig:bis78_gassystem}).  
Gas leak rates at 3~bar pressure below the limit of $2n \cdot 10^{-8}$~bar~l/s required for a chamber with n tubes 
have been achieved for all BMG chambers~\cite{BMGconstr}.

\section{sMDT chamber electronics}
After the installation of the gas distribution system, ground pins and Faraday cages, the high-voltage and the 
signal distribution boards (see figure~\ref{fig:sMDT_electronics}) as well as the active readout electronics (mezzanine) cards with 6 x 4 channels matching the transverse 
cross section of the quadruple-multilayers are mounted on opposite ends of the chambers.
The decoupling capacitors on the RO side and the terminating resistors on the HV side are enclosed in plastic
containers in order to guarantee HV stability. The mezzanine cards, stacked on top of the signal distribution boards, contain three 8-channel amplifier-shaper discriminator (ASD) 
chips and a TDC chip for drift time measurement which provide the same functionality as the standard MDT readout electronics.


\begin{thebibliography}{10}

\bibitem{ATLASpaper}
The ATLAS collaboration, G.~Aad et al., 
\emph{The ATLAS Experiment at the Large Hadron Collider},
\emph{J.Instr.} {\bf 3} (2008) S08003.

\bibitem{sMDTperf1}
B.~Bittner et al., 
\emph{Development of Muon Drift-Tube Detectors for High-Luminosity Upgrades of the Large Hadron Collider},
\emph{Nucl.~Instr.\ and Meth.} {\bf A617} (2010) 169.

\bibitem{MDTperf}
M.~Deile et al., 
\emph{Resolution and Efficiency of the ATLAS Muon Drift-Tube Chambers at High Background Rates},
\emph{Nucl.~Instr.\ and Meth.} {\bf A535} (2004) 212;
S.~Horvat et al., 
\emph{Operation of the ATLAS Muon Drift-Tube Chambers at High Background Rates and in Magnetic Fields},
\emph{IEEE Trans.~Nucl.~Sci.} {\bf 53, no.~2} (2006) 562.

\bibitem{sMDTprototype}
H.~Kroha et al., 
\emph{Construction and Test of a Full-Scale Prototype Drift-Tube Chamber for the Upgrade of the ATLAS Muon Spectrometer at High LHC Luminosities},
\emph{Nucl. Instr. and Meth.} {\bf A718} (2013) 427. 

\bibitem{BMEop}
C.~Ferretti, H.~Kroha (on behalf of the ATLAS Muon Collaboration), 
\emph{Upgrades of the ATLAS Muon Spectrometer With sMDT Chambers},
\emph{Nucl.~Instr.~and Meth.} {\bf A 824} (2016) 538.

\bibitem{BMGconstr}
H.~Kroha et al., 
\emph{Construction and Test of New Precision Drift Tube Chambers for the ATLAS Muon Spectrometer},
\emph{Nucl.~Instr.~and Meth.~A}, doi:10.1016/j.nima.2016.05.091.

\bibitem{BMGproduction}
H.~Kroha et al.,
\emph{Performance of New High-Precision Muon Tracking Detectors for the ATLAS Experiment},
arXiv:1701.08971, November 2016.

\bibitem{sMDTperf2}
B.~Bittner et al., 
\emph{Performance of Drift-Tube Detectors at High Counting Rates for High-Luminosity LHC Upgrades}, 
\emph{Nucl.~Instr.~Meth.} {\bf A732} (2013) 250.

\bibitem{blr} 
S.~Nowak et al.,  
\emph{Optimisation of the Read-out Electronics of Muon Drift-Tube Chambers for Very High Background Rates at HL-LHC and Future Colliders},
arXiv:1603.08841, November 2015.

\bibitem{aging}
O.~Kortner et al., 
\emph{Precision Muon Tracking Detectors and Read-Out Electronics for Operation at Very High Background Rates at Future Colliders},
\emph{Nucl.~Instr.~and Meth.} {\bf A824} (2016) 556.


\end{thebibliography}
\end{document}